\documentstyle[aps,preprint]{revtex}
\title{Disclination Asymmetry in 
Two-Dimensional Nematic Liquid Crystals with Unequal
Frank Constants}

\author{Michael W.\ Deem\\
Lyman Laboratory of Physics\\Harvard University, Cambridge, MA\ \ 02138}

\begin{document}
\bibliographystyle{prsty} 
\maketitle
\renewcommand{\baselinestretch}{1.3}

\begin{abstract}
The behavior of a thin film of nematic liquid crystal
with unequal Frank constants is discussed.  Distinct Frank
constants are found to imply unequal core energies for
$+1/2$ and $-1/2$ disclinations.  Even so, a topological 
constraint is shown to ensure that the bulk densities of the two types of
disclinations are the same.  For a system with free boundary conditions,
such as a liquid membrane, unequal core energies
simply renormalize the Gaussian rigidity and line tension.
\end{abstract}



\section{Introduction}
This paper discusses the disclination-mediated isotropic-ordered
transition in a thin film of nematic liquid crystal.  The focus is on the
case where the bend and splay Frank constants are distinct.  The
free energy of this system is given to lowest order by
\cite{Degennes}
\begin{equation}
H = \frac{k_1}{2} \int dx dy
\left[ \nabla \cdot {\bf n}(x,y) \right]^2  +
\frac{k_3}{2} \int dx dy
\left\vert \nabla \times {\bf n}(x,y) \right\vert^2 \ .
\label{1}
\end{equation}
Here $\bf n$ is the orientation of the nematic molecule and is of
unit length.  This free energy can alternatively be expressed in
terms of the orientation of the molecules as
\begin{eqnarray}
H &=& \frac{J}{2} \int dx dy \left( \theta_x^2 + \theta_y^2 \right)
+\frac{\Delta}{2} \int dx dy \cos(2 \theta) 
\left( \theta_y^2 - \theta_x^2 \right)
\nonumber \\
&&-\Delta \int dx dy \sin(2 \theta) 
 \theta_x \theta_y \ ,
\label{2}
\end{eqnarray}
where $J = (k_1 + k_3)/2$, $\Delta = (k_1-k_3)/2$, and the subscripts
denote derivatives.  If the two Frank constants were equal, this free
energy would simply be that of the X-Y model.  
For the symmetric nematics considered here, however, the natural defects are
$\pm 1/2$ disclinations, rather than the $\pm 1$ disclinations of the
conventional X-Y model.  It will be shown that the presence of non-zero
$\Delta$ causes the ground state energies of $\pm 1/2$ disclinations
to differ by $O(\Delta^2)$.  The disclination energies diverge
logarithmically in the system size, but with unequal coefficients.
The elementary Kosterlitz-Thouless energy-entropy balance thus seems
to lead to different proliferation temperatures for these defects.
This famous argument \cite{Kosterlitz} predicts that a $+1/2$ or $-1/2$
defect proliferates whenever the free energy to create a
disclination, $F_{+1/2}(R) = E_{+1/2}(R) - 2 k_{\rm B} T \ln(R/a_0)$
or $F_{-1/2}(R) = E_{-1/2}(R) - 2 k_{\rm B} T \ln(R/a_0)$, becomes
negative.  Here $E_{+1/2}(R)$ and $E_{-1/2} (R)$  are disclination
energies as a function of the system size $R$, and $a_0$ is a
microscopic cutoff.

In fact, thermal fluctuations of the nematics drive the two Frank
constants to the same value at long wavelengths, so
that there is a unique Kosterlitz-Thouless transition temperature. 
The essential effect of non-zero $\Delta$ is to create a distinct long-ranged
contribution to the core energy of each defect.  For a system above
the isotropic-ordered transition, this difference in core energies can be substantial
and would seem to lead to different
 densities of $+1/2$ and $-1/2$ disclinations.
As the correlation length grows near the isotropic-ordered
transition, the core energy becomes negligible compared to the
logarithmically diverging piece, and the disclinations pair into dislocations.
At the transition, the densities of $+1/2$ and
$-1/2$ disclinations must both become equal to the density of
dislocations.  In fact, Green's theorem implies that under
all conditions the difference
between the number of $+1/2$ and $-1/2$ disclinations  can scale at most as
the circumference of the system.  The natural way for this to happen is 
$n_{+1/2} - n_{-1/2} \sim c R/\xi$,
 where $\xi$ is the correlation length, and the prefactor depends
on difference of exponentials of core energies.
This constraint implies that $\pm 1/2$ disclinations occur with the same
density in a large system.

 The ground state energies of $\pm 1/2$ 
disclinations
are derived in Sec.\ II.  The energies are found to be
logarithmically diverging with unequal
prefactors.  It is shown in Sec.\ III
that these coefficients should renormalize 
to the same value at long wavelengths due to renormalization of
$\Delta$ to zero.  The free energies
of $\pm 1/2$ disclinations in the ordered phase are
directly calculated by perturbation
theory in Sec.\ IV, and  the $\pm 1/2$ disclinations are found to differ
by a core energy contribution.  An approximate
calculation of the disclination density in the isotropic phase is
described in Sec.\ V. 
The $\pm 1/2$
disclinations are shown to occur with equal densities for large systems.
In fact, the number asymmetry is shown to scale only linearly with the
system size.  This issue is explored in
Sec.\ VI with Monte Carlo calculations on a lattice model.
The disclination number asymmetry is indeed found to be
proportional to the circumference of the system.
Section VII concludes with a discussion of these results.


\section{Ground State Energies of $\pm 1/2$ Disclinations}
The ground state energies of $\pm 1/2$
disclinations in the Hamiltonian (\ref{2}) will be
logarithmically diverging in the size of the system.
The coefficient in front of the logarithm is calculated in this
section.  The
coefficient is determined solely by the properties of the $\theta$
field far from the disclination, vanishing if the $\theta$ field
vanishes at infinity and diverging unless the $\theta$ field goes to
a constant.  The ground state configuration, therefore, must have
$\theta(r, \phi) \sim \theta(\phi)$ as $r \to \infty$.  With this
form, the energy is given by
\begin{equation}
H \sim H_0 \ln(R/a_0) ~~{\rm as}~ R \to \infty \ ,
\label{3}
\end{equation}
with
\begin{equation}
H_0 = \frac{1}{2} \int_0^{2 \pi} d\phi~ {\theta'}^2(\phi) \left\{
J + \Delta \cos[2 \phi - 2 \theta(\phi)]
\right\} \ .
\label{4}
\end{equation}
A disclination of strength $s$ is defined by
\begin{equation}
\theta(\phi) = s \phi + \frac{\Delta}{J} \theta_1(\phi) \ ,
\label{5}
\end{equation}
where $\theta_1(\phi)$ is continuous.  The condition of a ground
state geometry, without the assumption of rotational symmetry,
 can be written as
\begin{eqnarray}
\frac{\delta F_0}{\delta \theta} = 0 &=& - J \nabla^2 \theta + 
\Delta \sin(2 \theta) (\theta_y^2 - \theta_x^2 + 2 \theta_{xy})
\nonumber \\
&&+\Delta \cos(2 \theta) (\theta_{xx} - \theta_{yy} + 2 \theta_x \theta_y) \ .
\label{5a}
\end{eqnarray}
This equation implies that 
\begin{equation}
\theta = \theta_0 + \frac{\Delta}{J} \theta_1 + O(\Delta^2/J^2) \ ,
\label{5b}
\end{equation}
with
\begin{eqnarray}
\nabla^2 \theta_0 &=& 0 \nonumber \\
\nabla^2 \theta_1 &=& \sin(2 \theta_0)
(\theta_{0y}^2 - \theta_{0x}^2 + 2 \theta_{0 xy})
\nonumber \\
&&+ \cos(2 \theta_0)
(\theta_{0xx} - \theta_{0yy} + 2 \theta_{0x}\theta_{0y})
\ .
\label{5bb}
\end{eqnarray}
These equations are solved by
\begin{eqnarray}
\theta_0(\phi) &=& s \phi
\nonumber \\
\theta_1(\phi) &=& \frac{(2 - s) s \sin[2 \phi(s-1)]}{4 (1-s^2)}
\label{5c}
\end{eqnarray}
for a disclination of strength $s$.

For $s = +1/2$, Eq.\ (\ref{5c}) simplifies to 
\begin{equation}
\theta_1(\phi) = - \frac{3 }{4 } \sin \phi
\label{6}
\end{equation}
and leads to an energy of
\begin{equation}
H_0  = \frac{\pi J}{4} - \frac{9 \pi \Delta^2}{32 J} + O(\Delta^3/J^2) \ .
\label{7}
\end{equation}
Similarly, for $s = -1/2$, Eq.\ (\ref{5c}) simplifies to
\begin{equation}
\theta_1(\phi) = \frac{5 }{36 } \sin(3 \phi)
\label{8}
\end{equation}
 and leads to an energy of
\begin{equation}
H_0  = \frac{\pi J}{4} - \frac{25 \pi \Delta^2}{288 J} + O(\Delta^3/J^2)
\ .
\label{9}
\end{equation}

The general result for the deviation field $\theta-\theta_0$ can be
expressed in terms of elliptic integrals of the third kind
\cite{Dzyaloshinski}.  Results for the angle field have been
presented elsewhere \cite{Hudson}.  To give the reader some feel for
how larger values of $\Delta$ distort the ground state geometry,
the function $\theta(\phi)$ is shown for a $s = +1/2$ (Fig.\ 1) and a $s =
-1/2$ (Fig.\ 2) disclination.  These geometries were computed by
defining $\theta(\phi)$ on a grid and numerically minimizing Eq.\
(\ref{4}).  Direct integration of Eq.\ (\ref{5a}) produced identical
results.  Extreme differences between the two Frank constants can
substantially distort the $T=0$ geometries.  Figure 3 shows the
ground state energies associated with different ratios of the Frank
constants for $s = \pm 1/2$.  One can see that the $s=+1/2$ disclination
completely screens out either splay or bend as the associated
Frank constant, $k_1$ or $k_3$, respectively, becomes large:
\begin{eqnarray}
H_0 \sim \pi k_3/2 ~{\rm as}~ k_1 \to \infty \nonumber \\
H_0 \sim \pi k_1/2 ~{\rm as}~ k_3 \to \infty
\label{10}
\end{eqnarray}
The $s=-1/2$ disclination, on the other hand, is unable to completely
remove unfavorable bend or splay:
\begin{eqnarray}
H_0 \sim 0.191 k_1 ~{\rm as}~ k_1 \to \infty \nonumber \\
H_0 \sim 0.191 k_3 ~{\rm as}~ k_3 \to \infty
\label{10aa}
\end{eqnarray}


\section{RG Flow equation for $\Delta$}
The isotropic-ordered transition does not occur at $T=0$, and so it
is the free energies of the two disclinations that should govern their
densities in the isotropic phase.  This
issue is addressed here by looking at the renormalization of $\Delta$
 due to thermal fluctuations  around an isolated disclination.

For a $s=+1/2$ disclination the angle order parameter is expressed as
\begin{equation}
\theta(r,\phi) = \phi/2 + \psi(r,\phi) \ .
\label{10a}
\end{equation}
The function $\psi(r,\phi)$ is single-valued and smooth.   
The Hamiltonian (\ref{2}) is
expanded in powers of $\psi$, with the result
\begin{eqnarray}
H[\psi] &=& \frac{J}{2} \int \left( \frac{1}{4 r^2} + \psi_x^2 + \psi_y^2
\right) + \frac{3 \Delta}{4} \int \frac{y \psi}{r^3} 
\nonumber \\
&&+ \frac{\Delta}{4} \int \left[
-\frac{3 x \psi^2}{r^3} +  \frac{2 x (\psi_y^2 - \psi_x^2)}{r} -
\frac{4 y \psi_x \psi_y}{r}
\right] \nonumber \\
&&+ \frac{\Delta}{6} \int \left[
-\frac{3 y \psi^3}{r^3} - \frac{6 y \psi(\psi_y^2 - \psi_x^2)}{r}
- \frac{12 x \psi \psi_x \psi_y}{r}
\right] \nonumber \\
&&+ \frac{\Delta}{12} \int \left[
\frac{3 x \psi^4}{r^3} - \frac{12 y \psi^2 (\psi_y^2 - \psi_x^2)}{r}
+ \frac{24 x \psi^2 \psi_x \psi_y}{r}
\right] \ .
\nonumber \\
\label{11} 
\end{eqnarray}
The renormalization of $\Delta$ is tracked by integrating out $\psi$ on
a momentum shell to first order in $T/J$ and watching how terms from
the $\psi^3$ expression contribute to  terms in the $\psi$
expression.  The $\psi$ expression can be written as
\begin{equation}
t_0 = \frac{3 \pi i \Delta}{2} \int_{\bf k} \frac{k_y}{k}
\hat \psi(-{\bf k}) \ ,
\label{12}
\end{equation}
where $\int_{\bf k}$ means $\int d^2 k / (2 \pi)^2$.
The $\psi^3$ expression can be broken down into 
\begin{eqnarray}
t_1 &=& -\pi i \Delta 
 \int_{{\bf k}_1 {\bf k}_2 {\bf k}_3 {\bf k}_4}
(2 \pi)^2 \delta({\bf k}_1 + {\bf k}_2 + {\bf k}_3 + {\bf k}_4)
\frac{k_{1 y}}{k_1}
\nonumber \\ &&~~~~~\times
 \hat \psi({\bf k}_2) \hat \psi({\bf k}_3) \hat \psi({\bf k}_4)
\nonumber \\
t_2 &=& - 2 \pi i \Delta
 \int_{{\bf k}_1 {\bf k}_2 {\bf k}_3 {\bf k}_4}
(2 \pi)^2 \delta({\bf k}_1 + {\bf k}_2 + {\bf k}_3 + {\bf k}_4)
\frac{k_{1 y}}{k_1^3}
\nonumber \\ &&~~~~~\times
 \hat \psi({\bf k}_2) \hat \psi({\bf k}_3) \hat \psi({\bf k}_4)
(k_{3 x} k_{4 x} - k_{3 y} k_{4 y})
\nonumber \\
t_3 &=& - 4 \pi i \Delta
 \int_{{\bf k}_1 {\bf k}_2 {\bf k}_3 {\bf k}_4}
(2 \pi)^2 \delta({\bf k}_1 + {\bf k}_2 + {\bf k}_3 + {\bf k}_4)
\frac{k_{1 x}}{k_1^3}
\nonumber \\ &&~~~~~\times
 \hat \psi({\bf k}_2) \hat \psi({\bf k}_3) \hat \psi({\bf k}_4)
(- k_{3 x} k_{4 y}) \ .
\label{13} 
\end{eqnarray}
The momenta in the shell $ k_{\rm c} -d k_{\rm c} < k < k_{\rm c}$
can then be integrated over with the result
\begin{eqnarray}
\left\langle t_1 \right\rangle_l &=&
-\pi i \Delta  \int_{\bf k} \frac{k_y}{k}
\hat \psi(-{\bf k}) \frac{6 \pi T k_{\rm c} d k_{\rm c}}
{J (2 \pi)^2 k_{\rm c}^2}
\nonumber \\
\left\langle t_2 \right\rangle_l &=& 0
\nonumber \\
\left\langle t_3 \right\rangle_l &=& 0 \ .
\label{14} 
\end{eqnarray}
This result implies
\begin{equation}
\Delta' = \Delta - \frac{\Delta T}{\pi J} \frac{d k_{\rm c}}{k_{\rm c}} \ .
\label{15}
\end{equation}
Defining $d k_{\rm c} / k_{\rm c} = dl$, the flow equation results:
\begin{equation}
\frac{d \Delta}{d l} = -\frac{\Delta T}{\pi J} \ .
\label{16}
\end{equation}
$J$ is not renormalized to $O(\Delta)$.

The same calculation can be performed for a $s=-1/2$ disclination, 
using the relation
\begin{equation}
\theta(r,\phi) = -\phi/2 + \psi(r,\phi) \ .
\label{16b}
\end{equation}
 The Hamiltonian (\ref{2}) is expanded in powers of $\psi$, with the result
\begin{eqnarray}
&&H[\psi] =
 \frac{J}{2} \int \left( \frac{1}{4 r^2} + \psi_x^2 + \psi_y^2
\right) + \frac{5 \Delta}{4} \int \frac{y^3 - 3 x^2 y}{r^5} \psi
\nonumber \\
&&+ \frac{\Delta}{4} \int \left[
\frac{5(x^3-3 x y^2) \psi^2}{r^5} +
  \frac{2 x (\psi_y^2 + \psi_x^2)}{r} +
\frac{4 y \psi_x \psi_y}{r}
\right] \nonumber \\
&&+ \frac{\Delta}{6} \int \left[
\frac{5 (3 x^2 y - y^3) \psi^3}{r^5} + \frac{6 y \psi 
   (\psi_y^2 - \psi_x^2)}{r}
- \frac{12 x \psi \psi_x \psi_y}{r}
\right] \nonumber \\
&&+ \frac{\Delta}{12} \int \left[
\frac{5 (3 x y^2 - x^3) \psi^4}{r^5} + \frac{12 x \psi^2
    (\psi_y^2 + \psi_x^2)}{r}
- \frac{24 y \psi^2 \psi_x \psi_y}{r}
\right]  \ .
\nonumber \\
\label{17} 
\end{eqnarray}
The renormalization of $\Delta$ is again tracked by integrating out
$\psi$ on a momentum shell to first order in $T/J$ and watching how
terms from the $\psi^3$ expression contribute to terms in the $\psi$
expression.  The flow equation that results is
\begin{equation}
\frac{d \Delta}{d l} = -\frac{\Delta T}{\pi J} \ .
\label{18}
\end{equation}
One can see that $\Delta$ renormalizes in the same way about $\pm 1/2$
disclinations.  In fact, the same flow equation describes the
renormalization of $\Delta$ in the absence of disclinations
\cite{Nelson}.

The renormalization of coupling, $J$, and the disclination
fugacity, $y$, can be studied with the correlation function approach used for the
standard X-Y model \cite{NelsonII}.  The result, combined with
the results above, is
\begin{eqnarray}
\frac {d T/J}{d l} &=& \pi^3 (y a_0^2)^2 + O(\Delta^2, y^4, y^2 \Delta)
\nonumber \\
\frac {d \Delta}{d l} &=& - \Delta T/(\pi J) + O(\Delta^2, y^2 \Delta)
\nonumber \\
\frac {d y a_0^2}{d l} &=& \left( 2 - \frac{\pi J}{4 T} \right) 
y a_0^2 + O(y^2 \Delta, y^3)
\ .
\label{19} 
\end{eqnarray}
One can see that at the critical point the renormalized
coupling is $J_{\rm R}/T = 8 / \pi$. 
Furthermore, $\Delta$ scales as
\begin{equation}
\Delta_{\rm R}^2 \sim \Delta^2 (\xi / a_0)^{-1/4} \ ,
\label{19a}
\end{equation}
where the correlation length is given by $\xi = a_0 \exp(l)$.


\section{Disclination Free Energies in Ordered Phase}
While the RG calculations show how $\Delta$ becomes irrelevant at the
isotropic-ordered transition, they do not directly show how the
disclination densities scale near the transition.  To estimate the
$\pm 1/2$ disclination densities, expressions are needed for the free
energies of isolated disclinations.
Perturbation theory is here used to calculate directly 
these free energies in the ordered phase.  The results should also be
applicable to correlated regions within a macroscopically
disordered phase near the isotropic-ordered transition.

The free energy of a disclination will be evaluated to $O(\Delta^2)$.
Equation (\ref{2}) is first integrated by parts with the result
\begin{eqnarray}
H &=& \frac{J}{2} \int dx dy \left( \theta_x^2 + \theta_y^2 \right)
-\frac{\Delta}{4} \int dx dy \sin(2 \theta) 
\left( \theta_{yy} - \theta_{xx} \right)
\nonumber \\
&&-\frac{\Delta}{2} \int dx dy \cos(2 \theta) 
 \theta_{xy} \ .
\label{20}
\end{eqnarray}
Although $\theta$ is discontinuous in the presence of a disclination,
the result is the same if Eq.\ (\ref{10a}) or (\ref{16b}) is used and
the integration by parts done in terms of $\psi$.  A
cumulant expansion is used for the free energy:
\begin{equation}
F = E_0 + \left\langle \delta H \right\rangle_{0{\rm c}} - \frac{1}{2}
\left\langle (\delta H)^2 \right\rangle_{0{\rm c}}/T + \ldots \ .
\label{21}
\end{equation}
Here the averages are done with respect to the reference system
with Hamiltonian $H_0$,
indicated by subscript zero, and are connected, indicated by
subscript c.  The functional $\delta H$ is $H - H_0$.  The
reference system is chosen to be
\begin{equation}
H_0[\psi] = \frac{J}{2} \int d {\bf r}
\left( \frac{1}{4 r^2} + \psi_x^2 + \psi_y^2
\right) \ ,
\label{22}
\end{equation}
where $\theta = s \phi + \psi$.

Specializing to the case of a $s = +1/2$ disclination, the perturbation
becomes
\begin{eqnarray}
\delta H[\psi] &=& \int d {\bf r} \bigg[
 \frac{\Delta y}{4 r^3} \sin(2 \psi) 
-\frac{\Delta y}{4 r} \cos(2 \psi)(\psi_{yy} - \psi_{xx})
 \nonumber \\ &&
-\frac{\Delta x}{2 r} \cos(2 \psi)\psi_{xy}
+ \frac{\Delta x}{4 r^3} \cos(2 \psi)
 \nonumber \\ &&
- \frac{\Delta x}{4 r} \sin(2 \psi)(\psi_{yy} - \psi_{xx})
+ \frac{\Delta y}{2 r} \sin(2 \psi)\psi_{xy}
\bigg] \ .
\nonumber \\
\label{23} 
\end{eqnarray}
A short calculation shows
\begin{equation}
\left\langle \delta H \right\rangle_{0{\rm c}} = 0 \ .
\label{24}
\end{equation}
The first non-zero contribution to the free energy is, therefore,
$O(\Delta^2)$.  The form of Eq.\ (\ref{23}), with three terms even in
$\psi$ and three terms odd, simplifies the evaluation of
$\left\langle (\delta H)^2 \right\rangle_{0{\rm c}}$.  Even so, there are
twelve Gaussian averages that must be performed.
A typical term is
\begin{eqnarray}
&&\int d {\bf r}_1 d {\bf r}_2
\left\langle 
\cos[2 \psi({\bf r}_1)] \psi_{xx}({\bf r}_1)
\cos[2 \psi({\bf r}_2)] \psi_{xy}({\bf r}_2)
\right\rangle_{0c} \nonumber \\
&=& 2 \int d {\bf r}_1 d {\bf r}_2
 e^{4 \chi(r_{12}) - 4 \chi(0)} 
\nonumber \\ && \times
\bigg\{[\chi_{xx}(r_{12}) - \chi_{xx}(0)]
[\chi_{xy}(r_{12}) - \chi_{xy}(0)]
 + \frac{1}{4} \chi_{xxxy}(r_{12})
\bigg\} \nonumber \\
&+&
2 \int d {\bf r}_1 d {\bf r}_2
 e^{-4 \chi(r_{12}) - 4 \chi(0)} 
\nonumber \\ && \times
\bigg\{ -[\chi_{xx}(r_{12}) + \chi_{xx}(0)]
[\chi_{xy}(r_{12}) + \chi_{xy}(0)] 
 + \frac{1}{4} \chi_{xxxy}(r_{12})
\bigg\} \ ,
\nonumber \\
\label{25} 
\end{eqnarray}
where $r_{12} = \vert {\bf r}_1 - {\bf r}_2 \vert$, and
\begin{eqnarray}
\hat \chi(k) &=& \frac{T}{J k^2} \nonumber \\
\chi(r) &=& \frac{T}{2 \pi J} \ln(R/r) \nonumber \\
\chi(0) &=& \frac{T}{2 \pi J} \ln(R/a_0)
\ .
\label{26} 
\end{eqnarray}
With the definition $z = 2 T / (\pi J)$, the first integral in Eq.\ (\ref{25})
scales like $(R/a_0)^{-z}$, and the second term scales like
$(R/a_0)^{-2 z} (R/a_0)^{z}$  Both terms must in principle be
evaluated.  However all twelve terms that contain the factor $
e^{-4 \chi(r_{12}) - 4 \chi(0)}$ cancel by $x
\leftrightarrow y$ symmetry.  The symmetry ${\bf r}_1
\leftrightarrow {\bf r}_2$ is applied to the other twelve terms with the
result
\begin{eqnarray}
 &&\int d {\bf r}_1 d {\bf r}_2
\left\langle (\delta H)^2 \right\rangle_{0{\rm c}}
= 
\nonumber \\ &&+
\frac{\Delta^2}{16} 
\int d {\bf r}_1 d{\bf r}_2 e^{4 \chi(r_{12})} \bigg[
\left\{ \frac{y_1}{r_1^3} \frac{y_2}{r_2^3} \right\}
\nonumber \\ && +
\left\{ 4 \frac{y_1}{r_1} \frac{y_2}{r_2}
[ \chi_{yy}(r_{12}) - \chi_{xx}(r_{12})]^2
 + 16 \frac{x_1}{r_1} \frac{x_2}{r_2}
 \chi_{xy}^2(r_{12})
 \right\} 
\nonumber \\ &&
+
\left\{ \frac{y_1}{r_1} \frac{y_2}{r_2} \nabla^4 \chi(r_{12})
\right\}
\nonumber \\ &&+
\left\{ - 4 \frac{y_1}{r_1^3} \frac{y_2}{r_2}
[ \chi_{yy}(r_{12}) - \chi_{xx}(r_{12})] -
8 \frac{y_1}{r_1^3} \frac{x_2}{r_2}
 \chi_{xy}(r_{12})
 \right\} 
\bigg] \nonumber \\
&& \equiv  I_1 + I_2 + I_3 + I_4
\label{27} 
\end{eqnarray}
with the redefinition $\chi(r_{12}) = -(z/4) \ln(r_{12}/a_0)$.  The four
integrals, $I_1$---$I_4$, represent integration over the four terms in braces.

These integrals can now be evaluated. This will be done in
Fourier space, and the following Fourier transforms will be helpful
\begin{eqnarray}
FT \left\{ r^{-z} \right\} &=& 2 \pi k^{z-2} 
    \frac{2 \Gamma(1-z/2) }{2^z \Gamma(z/2)}
\nonumber \\
FT \left\{y/r^3 \right\}&=& 2 \pi i \sin \phi
\nonumber \\
FT \left\{y/r \right\}&=& 2 \pi i \sin \phi / k^2
\label{28} \ .
\end{eqnarray}
The first of these transforms is well-defined for $0 < z < 2$.
The relation is valid for all $z$ by analytic continuation from the
relation $FT \left\{\nabla^2 f \right\} = -k^2 \hat f(k)$.  With these
definitions, the first integral becomes
\begin{eqnarray}
I_1 &=& \frac{\Delta^2 a_0^z}{16} \int_{\bf k}
\vert 2 \pi i \sin \phi \vert^2
 2 \pi k^{z-2} 
\frac{2 \Gamma(1-z/2) }{2^z \Gamma(z/2)}
\nonumber \\
&=& \frac{\Delta^2 \pi^2}{8} 
\frac{2 \Gamma(1-z/2) }{2^z \Gamma(z/2)}
 \frac{1 - (R/a_0)^{-z}}{z}
\nonumber \\
&=& \frac{\Delta^2 \pi^2 z}{8}
\frac{\Gamma(1-z/2) (1+z/2)}{2^z \Gamma(2+z/2)}
 \frac{1 - (R/a_0)^{-z}}{z}
 \ .
\label{29}
\end{eqnarray}
The second integral becomes
\begin{eqnarray}
I_2 &=&
 \frac{\Delta^2 z^2 }{16}
\int d {\bf r}_1 d{\bf r}_2 \left( \frac{r_{12}}{a_0}\right)^{-z-4} 
\frac{y_1}{r_1} \frac{y_2}{r_2}
\nonumber \\
&=&
 \frac{\Delta^2 z^2 a_0^z}{16} \int_{\bf k}
\left\vert \frac{2 \pi i \sin \phi}{k^2} \right\vert^2
  2 \pi k^{z+2} 
\frac{2 \Gamma(-1-z/2) }{2^{4+z} \Gamma(2+z/2)}
\nonumber \\
&=& \frac{\Delta^2 \pi^2 z^2}{8} 
\frac{2 \Gamma(-1-z/2) }{2^{4+z} \Gamma(2+z/2)}
 \frac{1 - (R/a_0)^{-z}}{z}
\nonumber \\
&=& \frac{\Delta^2 \pi^2 z}{32}
\frac{\Gamma(1-z/2) }{2^z \Gamma(2+z/2)(1+z/2)}
 \frac{1 - (R/a_0)^{-z}}{z}
\label{30} \ .
\end{eqnarray}
The following identity will be useful in evaluating
the third integral:
\begin{equation}
\chi(r) = -\frac{z}{4} \lim_{\alpha \to 0}
 \frac{1 - (r/a_0)^{-\alpha}}{\alpha} \ ,
\label{31}
\end{equation}
which implies
\begin{equation}
\nabla^4 \chi(r) = \frac{z}{4} \lim_{\alpha \to 0}
\alpha(\alpha+2)^2 (r/a_0)^{-\alpha-4} \ .
\label{32}
\end{equation}
Using this result, one finds
\begin{eqnarray}
z^{-1} FT \left\{ r^{-z} \nabla^4 \chi \right\} &=& 
\frac{1}{4} \lim_{\alpha \to 0}
\alpha(\alpha+2)^2
 \frac{k^4}{(z+\alpha)^2 (z + \alpha + 2)^2}
\nonumber \\ &&\times
\frac {4 \pi k^{z+\alpha  - 2} \Gamma[1 - (z+\alpha)/2]}
{2 ^{z+\alpha} \Gamma[(z+\alpha)/2]} \nonumber \\
&=&
\left\{
\begin{array}{l}
\pi k^2 /2,~ z=0 \\[.2in]
0, z \ne 0
\end{array}  \right.
 \ .
\label{33}
\end{eqnarray}
The term $I_3$, therefore, vanishes for non-zero temperatures.
It is convenient to break the fourth integral into two parts:
\begin{eqnarray}
I_{4a} &=&
  \frac{\Delta^2}{4}
\int d {\bf r}_1 d{\bf r}_2
\frac{y_1}{r_1^3} \frac{y_2}{r_2}
\left( \frac{r_{12}}{a_0} \right)^{-z}
 \left[ \chi_{xx}(r_{12}) - \chi_{yy}(r_{12}) \right]
\nonumber \\
I_{4b} &=&
 -\frac{\Delta^2}{2}
\int d {\bf r}_1 d{\bf r}_2
\frac{y_1}{r_1^3} \frac{x_2}{r_2}
\left( \frac{r_{12}}{a_0} \right)^{-z}
 \chi_{xy}(r_{12})
 \ .
\label{34}
\end{eqnarray}
The following trick is used to evaluate these terms:
\begin{eqnarray}
r^{-z} \left[ \chi_{xx}(r) - \chi_{yy}(r) \right]
&=& \lim_{\alpha \to 0} \frac{z}{4 \alpha} r^{-z}
(\partial_x^2 - \partial_y^2) r^{-\alpha}
\nonumber \\
&=& \frac{z}{2} r^{-z-2} \cos(2 \phi) \ .
\label{35}
\end{eqnarray}
In this form, the Fourier transforms can be evaluated, with the result
\begin{eqnarray}
I_{4a} &=& \frac{\Delta^2 a_0^z }{4} \int_{\bf k}
\frac{\vert 2 \pi i \sin \phi \vert^2}{k^2}
\left(  -\frac{\pi z}{2} \right) \cos(2 \phi) k^{z} 
\frac{\Gamma(1-z/2) }{2^{z} \Gamma(2+z/2)}
\nonumber \\
&=& \frac{\Delta^2 \pi^2 z}{16} 
\frac{\Gamma(1-z/2) }{2^{z} \Gamma(2+z/2)}
 \frac{1 - (R/a_0)^{-z}}{z} \ .
\label{36}
\end{eqnarray}
Similarly,
\begin{eqnarray}
I_{4b} &=& -\frac{\Delta^2 a_0^z }{2} \int_{\bf k}
(2 \pi i \sin \phi)^*
\frac{2 \pi i \cos \phi }{k^2}
\left( -\frac{\pi z}{4} \right)
\nonumber \\ &&\times
 \sin(2 \phi) k^{z} 
\frac{\Gamma(1-z/2) }{2^{z} \Gamma(2+z/2)}
\nonumber \\
&=& \frac{\Delta^2 \pi^2 z}{16} 
\frac{\Gamma(1-z/2) }{2^{z} \Gamma(2+z/2)}
 \frac{1 - (R/a_0)^{-z}}{z} \ .
\label{37}
\end{eqnarray}
Combining all these results one finds for the free energy of a $+1/2$
disclination at the origin
\begin{eqnarray}
F_{+1/2} &=& \frac{\pi J}{4} \ln(R/a_0) 
\nonumber \\ &&- \frac{\pi \Delta^2}{32 J}
\frac{\Gamma(1-z/2) }{2^z \Gamma(2+z/2)}
 \left( 8 + 2 z + \frac{1}{1 + z/2} \right)
\nonumber \\ && \times
 \frac{1 - (R/a_0)^{-z}}{z} + O(\Delta^3) \ .
\label{38}
\end{eqnarray}
The ground state energy is found to be
\begin{equation}
F_{+1/2} \sim \left( \frac{\pi J}{4} -
 \frac{9 \pi \Delta^2}{32 J} \right)
 \ln(R/a_0) + O(\Delta^3) ~{\rm as}~ T \to 0 \ ,
\label{39}
\end{equation}
in agreement with Eq.\ (\ref{7}).

For the case of 
a $s = -1/2$ disclination, the perturbation to consider is
\begin{eqnarray}
\delta H[\psi] &=& 
\frac{\Delta}{4}
\int d {\bf r} \bigg[
\frac{y^3 - 3 x^2 y}{r^5} \sin(2 \psi) 
+\frac{y}{r} \cos(2 \psi)(\psi_{yy} - \psi_{xx})
 \nonumber \\ 
&&-\frac{2 x}{r} \cos(2 \psi)\psi_{xy}
+ \frac{3 x y^2 - x^3}{r^5} \cos(2 \psi)
 \nonumber \\ &&-
 \frac{x}{r} \sin(2 \psi)(\psi_{yy} - \psi_{xx})
- \frac{2 y}{ r} \sin(2 \psi)\psi_{xy}
\bigg] \ .
\label{40} 
\end{eqnarray}
A short calculation shows
\begin{equation}
\left\langle \delta H \right\rangle_{0{\rm c}} = 0 \ ,
\label{41}
\end{equation}
so that the first non-zero contribution to the free energy is
$O(\Delta^2)$.  As before, averages that lead to terms with $ e^{-4
\chi(r_{12}) - 4 \chi(0)}$ cancel  by $x \leftrightarrow y$
symmetry.  Also as before, the term containing the $\nabla^4 \chi(r)$
vanishes at non-zero temperature.  After some simplification, one finds
\begin{eqnarray}
 \int d {\bf r}_1 d {\bf r}_2
\left\langle (\delta H)^2 \right\rangle_{0c}
&=&I_2 + \frac{\Delta^2}{16} 
\int d {\bf r}_1 d{\bf r}_2 e^{4 \chi(r_12)} \bigg[
\left\{ f_1 f_2 \right\} 
\nonumber \\ &&+
\left\{ 4 f_1 \frac{y_2}{r_2}
[ \chi_{yy}(r_{12}) - \chi_{xx}(r_{12})] \right\}
\nonumber \\ &&+
 \left\{ -
8 f_1 \frac{x_2}{r_2} \chi_{xy}(r_{12}) \right\}
\bigg] \ ,
\label{42} 
\end{eqnarray}
where $f_i = (y_i^3 - 3 x_i^2 y_i)/r_i^5$.  The result
\begin{equation}
\hat f({\bf k}) = i \pi
\left( -6 \sin \phi + 8 \cos^2 \phi \sin \phi
+ \frac{16}{3} \sin^3 \phi \right) 
\label{43}
\end{equation}
will be used.

The  integral (\ref{42}) is split into the three bracketed pieces.  The first
integral is
\begin{eqnarray}
I_5 &=& 
\frac{\Delta^2}{16} 
\int d {\bf r}_1 d{\bf r}_2 
\left( \frac{r_{12}}{a_0} \right)^{-z}
f_1 f_2 \nonumber \\
&=&
\frac{\Delta^2 a_0^z}{16} 
\int_{\bf k} \vert \hat f({\bf k}) \vert^2
2 \pi k^{z-2}
\frac{2 \Gamma(1-z/2) }{2^z \Gamma(z/2)}
\nonumber \\
&=&
 \frac{\Delta^2 \pi^2 z}{72}
\frac{\Gamma(1-z/2) (1+z/2)}{2^z \Gamma(2+z/2)}
 \frac{1 - (R/a_0)^{-z}}{z}
 \ .
\label{44}
\end{eqnarray}
The second integral is
\begin{eqnarray}
I_6 &=& 
\frac{\Delta^2}{16} 
\int d {\bf r}_1 d{\bf r}_2 
\left( \frac{r_{12}}{a_0} \right)^{-z}
4 f_1 \frac{y_2}{r_2}
[ \chi_{yy}(r_{12}) - \chi_{xx}(r_{12}) ] 
 \nonumber \\
&=&
\frac{\Delta^2 a_0^z}{4} 
\int_{\bf k} \hat f^*({\bf k})
\frac{2 \pi i \sin \phi }{k^2}
  \frac{\pi z}{2} \cos(2 \phi) k^{z} 
\frac{\Gamma(1-z/2) }{2^{z} \Gamma(2+z/2)}
\nonumber \\
&=&
 \frac{\Delta^2 \pi^2 z}{48}
\frac{\Gamma(1-z/2)}{2^z \Gamma(2+z/2)}
 \frac{1 - (R/a_0)^{-z}}{z}
 \ .
\label{45}
\end{eqnarray}
The third integral is
\begin{eqnarray}
I_7 &=& 
\frac{\Delta^2}{16} 
\int d {\bf r}_1 d{\bf r}_2 
\left( \frac{r_{12}}{a_0} \right)^{-z}
\left( -8 f_1 \frac{x_2}{r_2} \right)
 \chi_{xy}(r_{12})
 \nonumber \\
&=&
-\frac{\Delta^2 a_0^z}{2} 
\int_{\bf k} \hat f^*({\bf k})
\frac{2 \pi i \cos \phi }{k^2}
 \left( - \frac{\pi z}{4}\right)
\nonumber \\ && \times \sin(2 \phi) k^{z} 
\frac{\Gamma(1-z/2) }{2^{z} \Gamma(2+z/2)}
\nonumber \\
&=&
 \frac{\Delta^2 \pi^2 z}{48}
\frac{\Gamma(1-z/2)}{2^z \Gamma(2+z/2)}
 \frac{1 - (R/a_0)^{-z}}{z}
 \ .
\label{46}
\end{eqnarray}
Combining all these results one finds for the free energy of a $-1/2$
disclination at the origin
\begin{eqnarray}
F_{-1/2} &=& \frac{\pi J}{4} \ln(R/a_0) 
\nonumber \\ &-&
\frac{\pi \Delta^2}{288 J}
\frac{\Gamma(1-z/2) }{2^{z} \Gamma(2+z/2)}
 \left( 16 + 2 z + \frac{9}{1 + z/2} \right)
\nonumber \\ && \times
 \frac{1 - (R/a_0)^{-z}}{z} + O(\Delta^3)\ .
\label{47}
\end{eqnarray}
The ground state energy is given by
\begin{equation}
F_{-1/2} \sim \left( \frac{\pi J}{4} -
 \frac{25 \pi \Delta^2}{288 J} \right)
 \ln(R/a_0) +
 O(\Delta^3) ~{\rm as}~ T \to 0 \ ,
\label{48}
\end{equation}
in agreement with Eq.\ (\ref{9}).

The difference in free energies of $\pm 1/2$ disclinations
is, therefore, given by
\begin{eqnarray}
F_{+1/2} - F_{-1/2} &=&
\frac{\pi \Delta^2}{J}
\frac{\Gamma(1-z/2) }{2^{z} \Gamma(2+z/2)}
\frac{7 + 2 z}{36}
 \frac{1 - (R/a_0)^{-z}}{z} 
\nonumber \\ &+& O(\Delta^3)\ .
\label{49}
\end{eqnarray}
Near the isotropic-ordered transition, the coupling
renormalizes to $z_{\rm R} = 1/4$ and
$J_{\rm R}/T = 8 / \pi$, so one finds
\begin{eqnarray}
(F_{+1/2} - F_{+1/2})/T &\sim&
-0.8891 (\Delta/T)^2 [1 - (\xi/a_0)^{-1/4}] 
\nonumber \\ &&+
[E_{{\rm c}_{+1/2}}(\Delta) - E_{{\rm c}_{-1/2}}(\Delta)]/T 
\nonumber \\ &&
~~{\rm as}~~ \xi \to \infty \ .
\label{50}
\end{eqnarray}
Additional microscopic core energies that may be
distinct for the two different disclinations have been explicitly
added in this equation.  By comparing Eq.\
(\ref{50}) with  Eqs.\ (\ref{7}), (\ref{9}), and (\ref{19a}), one sees
that the appropriate finite-size scaling replacement is $\Delta_{\rm R}^2
\ln(R/a_0) \to \Delta^2   [1 - (R/a_0)^{-z}]/z$.  This relation gives
a very good approximation to Eq.\ (\ref{50}).


\section{Disclination Densities in Isotropic Phase}
The disclination density  in
the isotropic phase is calculated in this section,
 taking into account interactions between
disclinations.  In the isotropic phase, the free energy can be
expressed in terms of the order parameter
\cite{Mermin}
\begin{eqnarray}
Q  &=& Q_0 
\left(
\begin{array}{c c}
n_x^2-n_y^2 & 2 n_x n_y \\[.2in]
2 n_x n_y & n_y^2 - n_x^2
\end{array}  \right)
\nonumber \\
  &=& Q_0 
\left(
\begin{array}{c c}
\cos (2 \theta) & \sin(2 \theta) \\[.2in]
\sin (2 \theta) & -\cos(2 \theta)
\end{array}  \right)
\nonumber \\
  &\equiv& 
\left(
\begin{array}{c c}
q_1 & q_2  \\[.2in]
q_2 & -q_1
\end{array}  \right) \ .
\label{51}
\end{eqnarray}
An expression that reduces to Eq.\ (\ref{2}) in the low-temperature,
fixed-$Q_0$ phase is
\begin{eqnarray}
H &=&  \frac{m}{4} \sum_{ij} \int d {\bf r} Q_{ij}^2 +
\frac{j_1}{4} \sum_{ijk} \int d {\bf r}
 \left( \partial_i Q_{jk} \right)^2
\nonumber \\ &+& 
 j_2 \sum_{ijst} \int d {\bf r}
Q_{si} Q_{tj}  \partial_s \partial_t Q_{ij} \ ,
\label{52}
\end{eqnarray}
where $j_1 = J/(4 Q_0^2)$ and
 $j_2 = \Delta/(16 Q_0^3)$.
Expressed in terms of the unique components $q_1$ and $q_2$, the
Hamiltonian becomes
\begin{equation}
H = 
\frac{m}{2} \int d {\bf r} (q_1^2 + q_2^2) +
\frac{j_1}{2} \int d {\bf r} (q_{1x}^2 + q_{1y}^2 +
q_{2x}^2 + q_{2y}^2) +
\delta H \ ,
\label{53}
\end{equation}
where
\begin{eqnarray}
\delta H = j_2 \int d {\bf r} \bigg[&&
q_1^2(q_{1xx} - q_{1yy}) 
+ q_2^2(q_{1yy} - q_{1xx}) 
\nonumber \\ && 
+ 2 q_1 q_2 (q_{2xx} - q_{2yy})  \nonumber \\&&
+ 4 q_1 q_2 q_{1xy}
+ 2 q_2^2 q_{2xy}
- 2 q_1^2 q_{2xy}
\bigg] \ .
\label{54}
\end{eqnarray}

The vector field ${\bf q} = (q_1, q_2)$ will have disclinations of
strength $\pm 1$ when the $\theta$ field has disclinations of
strength $\pm 1/2$.  The density of disclinations can be written as
\cite{Halperin}
\begin{equation}
\rho({\bf r}) = 
\frac{1}{2} \sum_l {\rm sgn} [\det \partial_i q_j({\bf r})]
\delta({\bf r} - {\bf r}_l) \ ,
\label{55}
\end{equation}
where ${\bf q}({\bf r}_l) = {\bf 0}$.  This expression can
be simplified as 
\begin{eqnarray}
\rho({\bf r}) &=& 
\frac{1}{2} \delta({\bf q}) \det \partial_i q_j({\bf r})
\nonumber \\
&=&
\frac{1}{2} \delta(q_1) \delta(q_2)
 (q_{1x} q_{2y} - q_{1y} q_{2x}) \ .
\label{56}
\end{eqnarray}
Furthermore, the number density is given by
\begin{equation}
\vert \rho({\bf r}) \vert = 
\frac{1}{2} \delta(q_1) \delta(q_2)
 \vert q_{1x} q_{2y} - q_{1y} q_{2x} \vert \ .
\label{57}
\end{equation}

These densities will be evaluated by perturbation theory in $\Delta$.
The reference system will be Eq.\ (\ref{53}), with $\delta H = 0$.
It is clear that $q_1$ and $q_2$ are independent Gaussian fields with
correlation function $\chi_0(r) = T / (2 \pi j_1) K_0(r/\xi)$,
where the correlation length is given by
$\xi= (j_1/m)^{1/2}$.  
Averages that involve the factor $\delta(q_1)$ will occur. This
factor can essentially be absorbed into the Gaussian weight with the
following trick:
\begin{eqnarray}
\left\langle  \delta(q({\bf 0})) f[q] \right\rangle_0 &=&
\frac{\int {\cal D}[q] P[q] \delta(q({\bf 0})) f[q]}
{\int {\cal D}[q] P[q]}
\nonumber \\
&=&
\frac{\int {\cal D}[q] P[q] \delta(q({\bf 0})) f[q]}
{\int {\cal D}[q] P[q] \delta(q({\bf 0})) }
\frac{\int {\cal D}[q] P[q] \delta(q({\bf 0}))}
{\int {\cal D}[q] P[q]}\ ,
\nonumber \\
\label{58}
\end{eqnarray}
where $P[q] {\cal D}[q]$
is the probability of a given field configuration.
This result is defined to be
\begin{equation}
\left\langle f[q]  \right\rangle_\delta 
\left\langle \delta(q({\bf 0})) \right\rangle_0
= [ 2 \pi \chi_0(0)]^{-1/2} 
\left\langle f[q]  \right\rangle_\delta \ .
\label{59}
\end{equation}
It turns out that the weight $P[q] \delta(q({\bf 0}))$ still implies that q
is a Gaussian field, but with the new correlation function
\cite{KardarI,KardarII,Chandler}
\begin{eqnarray}
\langle q({\bf r}_1) q({\bf r}_2)\rangle_\delta \equiv
\chi({\bf r}_1, {\bf r}_2) &=&
\chi_0(\vert {\bf r}_1 -{\bf r}_2 \vert) 
\nonumber \\ &-&
\chi_0(r_1) \chi_0(r_2) / \chi_0(0) \ .
\label{60}
\end{eqnarray}

The average number density of disclinations in the
limit $\Delta \to 0$ is first evaluated.  The following
term will arise
\begin{eqnarray}
\left\langle
\delta(q_1({\bf 0})) \delta(q_2({\bf 0})) 
\vert q_{1x}({\bf 0}) q_{2y}({\bf 0}) -
q_{1y}({\bf 0}) q_{2x}({\bf 0}) \vert \right\rangle_0 
\nonumber \\
= [2 \pi \chi_0(0)]^{-1} 
\left\langle
\vert q_{1x}({\bf 0}) q_{2y}({\bf 0}) -
q_{1y}({\bf 0}) q_{2x}({\bf 0}) \vert \right\rangle_\delta \ .
\label{61}
\end{eqnarray}
Using Eq.\ (\ref{60}), one can see that all of the variables in this average
are independently Gaussian, with variance $\gamma = 
-\chi_{0xx}(0) = -\chi_{0yy}(0) = -\chi_{0rr}(0)$.
  The probability distribution for $P(q_{1x} q_{2 y} = x)$
is $K_0(x/\gamma) / (\pi \gamma)$.  And  the probability
distribution for $P(q_{1 x} q_{2 y} - q_{1y} q_{2 x} = y)$ is
$\exp(-\vert y \vert/\gamma) / (2 \gamma)$.  With this result the
average can be carried out:
\begin{equation}
2 \langle \vert \rho \vert\rangle =
\frac{-\chi_{0rr}(0)}{2 \pi \chi_0(0)} + O(\Delta) \ .
\label{62}
\end{equation}

The effect of non-zero $\Delta$
on the average disclination asymmetry is now
evaluated.  From
Sec.\ IV, one knows that the essential effect of non-zero $\Delta$ is to
create distinct core energies for $+1/2$ and $-1/2$ disclinations.
This effect is modeled by replacing Eq.\ (\ref{54}) with
\begin{equation}
\delta H = - 2 \int d{\bf r} \mu(r) \rho({\bf r}) \ .
\label{63a}
\end{equation}
The parameter $\mu$ is related to the core energy difference, and
there will be a unique mapping from $\Delta^2$ to $\mu$
for small $\Delta$.  
The core energy difference is considered to be non-zero and constant within a
large region of radius $R$ inside a macroscopically large system.
  This is done because the macroscopic disclination
asymmetry is identically zero for a system with the periodic boundary
conditions implied by the Fourier analysis.
A grand canonical ensemble with a fluctuating disclination
asymmetry arises when considering a region of the periodic system.
 The disclination asymmetry is now calculated for
small $\mu$:
\begin{equation}
\langle \rho({\bf 0}) \rangle = 
\langle \rho({\bf 0}) \rangle_0  + 2\mu \int_{r < R} d {\bf r}
\langle \rho({\bf 0}) \rho({\bf r}) \rangle_{0 \rm c}   + O(\mu^2)\ .
\label{63}
\end{equation}

The disclination asymmetry is seen to be related to the
disclination correlation function by a fluctuation-dissipation
theorem.  The disclination correlation function is given by
\cite{Halperin}
\begin{equation}
4 \langle \rho({\bf 0}) \rho({\bf r}) \rangle_0 =
2 \langle \vert \rho \vert \rangle_0
+ g_0(r) \ ,
\label{64}
\end{equation}
where the delta function comes from the self terms in the average,
and $g_0(r)$ is a radially symmetric function that accounts for
correlations between distinct disclinations.
To calculate $g_0(r)$ the same
trick as before is used.  One has
\begin{equation}
\langle \delta(q({\bf 0})) \delta(q({\bf r})) \rangle_0 = \frac{1}
{2 \pi [\chi_0^2(0)  - \chi_0^2(r)]^{1/2}} \ .
\label{65}
\end{equation}
  The field correlation function is now given by
\cite{Chandler}
\begin{eqnarray}
\chi({\bf r}_1, {\bf r}_2) &=&
\chi_0(\vert {\bf r}_1 -{\bf r}_2 \vert) - \bigg[
\chi_0(r_1) \chi_0(r) \chi_0(r_2)
\nonumber \\ &&-
\chi_0(r_1) \chi_0(r) \chi_0(\vert{\bf r}_2-{\bf r}\vert)
\nonumber \\ &&-
\chi_0(\vert{\bf r}_1 - {\bf r} \vert ) \chi_0(r) \chi_0(r_2)
\nonumber \\ &&  +
\chi_0(\vert{\bf r}_1 - {\bf r} \vert ) \chi_0(r)
      \chi_0(\vert{\bf r}_2-{\bf r}\vert)
\bigg]
\nonumber \\ &&
\times [\chi_0^2(0)  - \chi_0^2(r)]^{-1}
\ .
\label{66}
\end{eqnarray}
This form implies
\begin{eqnarray}
\langle q_{1 \alpha}({\bf 0}) q_{1 \beta}({\bf r}) \rangle_\delta &=&
-\chi_{0 \alpha \beta}(r) 
\nonumber \\ &&-
\chi_{0 \alpha}(r) \chi_0(r) \chi_{0 \beta}(r) 
 / [\chi_0^2(0)  - \chi_0^2(r)] \ .
\label{67}
\end{eqnarray}
With these results one finds
\begin{equation}
g_0(r) = \frac{1}{2 \pi^2 r [\chi_0^2(0)  - \chi_0^2(r)]}
\left[
\chi_{0r}(r) \chi_{0rr}(r) + \frac{\chi_0(r) \chi_{0r}^3(r)}
{\chi_0^2(0)  - \chi_0^2(r)}
\right] \ .
\label{68}
\end{equation}
Per Eq.\ (\ref{63}) this result will be integrated over $r$.  
The first term can be integrated by parts, with the result 
\begin{equation}
\int d {\bf r} g_0(r) = \lim_{r \to 0} \frac{\chi_{0r}(r)}
{2 \pi r \chi_0(0)}  = -2\langle \vert \rho \vert \rangle_0 \ .
\label{69}
\end{equation}
This result, along with Eqs.\ (\ref{63}) and Eq.\ (\ref{64}),
implies that the densities of $+1/2$ and $-1/2$ disclinations
remain equal even in the presence of distinct core energies.
A more careful conclusion is that the average disclination density
is zero within the bulk region.  It can be non-zero 
near the boundary because the integral in Eq.\ (\ref{63})
will be cut off before $g_0(r)$ is negligible.  
This contribution is estimated to be
\begin{eqnarray}
\langle n_{+1/2} - n_{-1/2}\rangle  &=& 2 \mu 
\int_{r,r'<R} d {\bf r} d {\bf r}' \langle \rho({\bf r})
\rho ({\bf r}') \rangle_0 + O(\mu^2)
\nonumber \\
&\approx& 2 \pi \mu R \xi \langle \vert \rho \vert \rangle /2
\nonumber \\ 
&\sim&  c 
[\exp(-\beta E_{{\rm c}_{+1/2}}) - \exp(-\beta E_{{\rm c}_{-1/2}})]
\nonumber \\ && \times
R/\xi  ~~{\rm as}~~ \xi \ll R \to \infty
 \ .
\label{69a}
\end{eqnarray}
The scaling $\langle \vert \rho \vert \rangle \sim 
c \xi^{-2}$ has been used.  It arises arises because the disorder
created by unbound disclinations defines the correlation length.


\section{Monte Carlo Calculations on a Lattice Model}
This section describes both a lattice model for nematics with unequal
Frank constants and a Monte Carlo procedure for evaluating
the properties of the model.
The fundamental degrees of freedom are spins of unit
length on a square lattice.  The Hamiltonian has both nearest-
and next-nearest-neighbor couplings:
\begin{eqnarray}
H &=& 
\frac{J}{4} \sum_i \sum_{j<5}
   \left[1 - ({\bf n}_i \cdot {\bf n}_j)^2  \right]
\nonumber \\ &&+
\frac{\Delta}{8} \sum_i \sum_{j<5} (-1)^j
   \left[1 - ({\bf n}_i \cdot {\bf n}_j)^2  \right]
\left[ n_{ix}^2 + n_{jx}^2 - n_{iy}^2 - n_{jy}^2 \right]
\nonumber \\ &&+
\frac{\Delta}{8} \sum_i \sum_{4<j<9} (-1)^j
   \left[1 - ({\bf n}_i \cdot {\bf n}_j)^2  \right]
\left[ n_{ix}n_{iy} + n_{jx}n_{jy} \right]
\ .
\nonumber \\ 
\label{77}
\end{eqnarray}
The sum over $i$ is over sites on a square lattice.
The sum over $j$ is defined according to Fig.\ 4.   Terms
that would place $j$ off the lattice are ignored.  This Hamiltonian
satisfies the symmetry ${\bf n}_i \leftrightarrow -{\bf n}_i$.  The
couplings between spins are symmetric under the interchange 
${\bf n}_i \leftrightarrow {\bf n}_j$.
Finally, in the limit of a very small lattice spacing, Eq.\ (\ref{77})
reduces to Eq.\ (\ref{2}).  Numerical values of
Eq.\ (\ref{77}) agree with those from  Eqs.\ (\ref{3})-(\ref{4}) for
specific forms of the $\theta$ field.

This model will be equilibrated
with a simple Metropolis move that perturbs
individual spins.  Specifically, a vector
randomly distributed in a disk of radius $r$ is added to a randomly chosen
spin.  A 
value $r=10$ will be found to be satisfactory.  The new spin is then normalized
to unit length.  If the energy of the lattice is lowered by
using this new spin, the new spin is adopted.  Otherwise, the new
spin is adopted with a probability $\exp[(E_{\rm o}-E_{\rm n})/T]$.  This move
satisfies detailed balance, and so this Monte Carlo procedure will
sample the Boltzmann distribution \cite{Kalos}.  A natural unit of 
equilibration time, the Monte Carlo step (MCS), is $N^2$ iterations of
this move on a $N \times N$ lattice.  Most runs last
for 800000 MCS after an initial equilibration of 80000 MCS.
For the case of $\Delta=2$, 3200000 MCS will be performed after an
equilibration time of 320000 MCS.
The properties of the lattice will be sampled every 50 MCS.

The number of disclinations can be counted by looking at all
$(N-1)^2$ plaquettes on the lattice and determining
whether a disclination of strength $+1/2$, $-1/2$, or $0$ is present
at each plaquette.
This determination is made by first defining
${\bf n}_1$-${\bf n}_4$ to be a counterclockwise
ordering of the spins around the plaquette.  
Double-headed spins are converted into single-headed spins by $m_{ix} =
n_{ix}^2 - n_{iy}^2$ and $m_{iy} = 2 n_{ix} n_{iy}$.  The following
angle is defined for the plaquette:
\begin{equation}
t =  \theta_{21}
 + \theta_{32}
 + \theta_{43}
 + \theta_{14} \ ,
\label{80}
\end{equation}
where $\theta_{ij} = \theta_i - \theta_j$ is
constrained to be in the range $(-\pi, \pi)$, and $\theta_i$ is the
angle associated with spin ${\bf m}_i$.
The disclination density at this plaquette is defined by
\begin{eqnarray}
s = \frac{t}{4 \pi}
\label{81} \ .
\end{eqnarray}
The disclination asymmetry is equal to the sum of this density over
the entire lattice.  By the analog of Green's theorem, 
it can be written as a sum of $\theta_{ij}$ 
over the boundary of the lattice.

The correlation length and the correlation time will be measured
during a Monte Carlo run to monitor convergence.
 The position correlation function is
defined as
\begin{equation}
g({\bf r}) = N^{-2} \sum_{\bf x}
 \left\{ 
[{\bf n}({\bf x}) \cdot {\bf n}({\bf x}-{\bf r})]^2 
- 1/2 \right\} \ ,
\label{78}
\end{equation}
where the ${\bf r}$ and ${\bf x}$ are integer vectors on the $N \times
N$ square lattice.  Free boundary conditions imply that
there will be significant effects of the boundary in this correlation
function.  The correlation length is defined in terms of the
position correlation function by
\begin{equation}
\xi = \int d{\bf r} r g({\bf r}) / 
 \int d{\bf r} g({\bf r})  \ .
\label{79}
\end{equation}
Here $\vert {\bf r} \vert$ ranges from $0$ to $2^{1/2} N$.  Fast
Fourier transforms will be used to compute this quantity in $O(N^2 \ln^2 N)$ time
\cite{Press}.  The time correlation function  is defined as
\begin{equation}
f(t) = \sum_{t'} {\bf m}(t') \cdot {\bf m}(t'-t) \ ,
\label{80a}
\end{equation}
where the ${\bf m}(t) = \sum_{\bf x} {\bf m}({\bf x}, t)$ is the vector sum of the
single-headed spins at MCS $t$.  A correlation time could be defined
by analogy with Eq.\ (\ref{79}), but noise in $f(t)$ for large $t$ causes
this approach to be unsatisfactory.  Instead, the correlation time,
$\tau$, is definedby the smallest value of $\tau$ for which $f(\tau)/f(0) < 1/e$.

The interesting observable is the asymmetry between the number of
$+1/2$ and $-1/2$ disclinations.  This asymmetry will be small, and it
will be important to quantify the statistical error in this
observable. The
difference $\delta n = n_{+1/2} - n_{-1/2}$ must scale as
$N/\xi$ by Green's theorem.  The variance
of this observable should scale with the observable and
inversely with the number of independent samplings:
\begin{equation}
\sigma_{\delta n} = c \left[ \frac{N/ \xi}{T/\tau} \right]^{1/2} \ ,
\label{82}
\end{equation}
where $T$ is the total number of MCS.  This variance is
independent of $\Delta$ for small $\Delta$, and so
the constant $c$ can be determined from the case where $\Delta = 0$.
In this limit,
the values of $\tau$ and $\xi$ can also be determined from the case $\Delta = 0$.

Figure 5 shows the correlation length as a function of
temperature for various lattice sizes.  The correlation
length grows at lower temperatures.  It
saturates at a fraction of the lattice size for even lower values of the
temperature.  
The isotropic-ordered transition temperature can be identified by the  
inflection point of this curve.  Finite-size scaling can be used to
extrapolate the inflection point as a function of $1/N$ to $N \to
\infty$.  The result is approximately $J_{\rm R}/T = 4$.
Figure 6 shows the correlation time, in units of 50 MCS,
as a function of temperature for various lattice sizes.
This time also grows for lower temperatures and larger
lattices.  In fact, near the critical point, the expected
scaling $\tau \sim c \xi^2$ is observed.

Figure 7 shows the observed number of disclinations as a function
of temperature for various lattice sizes.  As expected,
the number of disclinations decreases with decreasing temperature.
  These curves
should converge to a universal curve in the limit $N \to \infty$.
This curve is the total number  of disclinations, both bound and
unbound, so it does not go to zero at the critical point.
The curves for different $N$, however, do intersect
at a unique value of $J/T$.  This value can be extrapolated
as a function of $1/N$ to $N \to \infty$ to determine
the true critical point.  The result is $J_{\rm R}/T = 3.4$.  This
observable appears to produce a more reliable critical point than does
the correlation length.

Not shown are the correlation length and disclination densities for
non-zero $\Delta$.  The dominant dependence
on $\Delta$ was through $\Delta^2$.  Non-zero values of $\Delta$
increased the density of disclinations and decreased the correlation
length and time.  Furthermore, analysis showed that the observed
disclination density scaled like $\rho \sim c \xi^{-2}$, at least for
large $T$, where the disclinations were unpaired.    These
qualitative effects are consistent with Eqs.\ (\ref{38}) and
(\ref{47}).

The limitation of a finite-size lattice prevents
the true scaling of the disclination asymmetry near
the isotropic-ordered transition from being observed.
  One can, however, observe the
asymmetry for these finite-size systems.  Figure 8 shows the
asymmetry function $\rho_{+1/2} - \rho_{-1/2}$ for $\Delta=2$ for various
lattices sizes.  Figure 9 shows the analogous
results for $\Delta=3$.  This observable can be extrapolated as a
function of $1/ N$ to $N \to \infty$ to estimate the behavior for
infinite systems.  It is clear that the extrapolated result is
zero, within statistical error.  In fact, one can see that the
disclination number asymmetry scales only with the circumference of
the system.

\section{Discussion}
We see that there is a statistical symmetry in this model.  This
symmetry enforces $\rho_{+1/2} + \rho_{-1/2}=0$ in the limit of 
a large system size, even in the presence of non-zero $\Delta$.
This symmetry arises because the disclination number asymmetry can
be written as an integral of bounded terms over the periphery of the
system, by Green's theorem.  This relation, in turn,
means that the disclination number
asymmetry can scale at most with the linear size of the system.
Indeed, this scaling was observed in the Monte Carlo calculations.
With a definition of disclinations not
susceptible to Green's theorem, this statistical symmetry would not
be present. In this case, one would generally expect the
disclination asymmetry to scale with $N^2/\xi^2$ instead of
$N/\xi$.

There are many parallels between the problem of a nematic liquid
crystal with unequal Frank constants and a hexatic membrane
\cite{Deem,ParkIII}.  Disclinations mediate a melting transition in both
cases, and the disclination free energy is logarithmic in the
correlation length in both cases.  For the
nematic, unequal Frank constants cause
the free energies of $+1/2$ and $-1/2$ disclinations
to differ.  They only
differ by a core energy, however, due to the renormalization
of $\Delta$.  For the membrane, local buckling causes the
the free energies of five- and
sevenfold disclinations to differ.  They, too, only differ by a core
energy, due to renormalization of the membrane rigidity.
As with nematic liquid crystals, there is a topological theorem
that relates the number of five- and sevenfold disclinations
to an integral over a boundary \cite{DavidII}.
The natural way for this to happen is
$n_5 - n_7 \sim c R/\xi_{\rm A}$, where $\xi_{\rm A}$ is now the hexatic
correlation length.  The prefactor again depends on the difference between
exponentials of core energies.  We conclude, then, that within the liquid
phase of a membrane with hexatic symmetry,
differing core energies simply renormalize the line tension and
Gaussian rigidity.  Near the liquid to hexatic
transition, however, constraints imposed by a non-Euclidean membrane
geometry frustrate the formation of hexatic
order.   The system can relieve this frustration by flattening
the membrane.  One might expect, for example, that the preferred size of a vesicle
undergoing a hexatic to liquid transition should scale like the hexatic
correlation length.  This conjecture is a worthy subject of future calculations.

\section*{Acknowledgements}
It is a pleasure to acknowledge discussions with David Nelson
and Georg Foltin.  This
research was supported by the National Science Foundation  through
Grant Nos.\ DMR--9417047 and CHE--9403114, and through the MRSEC
program through Grant No.\ DMR--9400396.

\bibliographystyle{new}

\section*{Figure Captions}

\flushleft{Figure 1.}
$\theta$ versus $\phi$ for a $s=+1/2$ disclination.

\flushleft{Figure 2.}
$\theta$ versus $\phi$ for a $s=-1/2$ disclination.

\flushleft{Figure 3.}
$E/J$ versus $\Delta$ for $s=+1/2$ (dashed line) and $s=-1/2$ (solid line).

\flushleft{Figure 4.}
Definition of spins used in Hamiltonian (\ref{77}).

\flushleft{Figure 5.}
The correlation length as a function of temperature
for the case $J=3$ and $\Delta =0$ and $N=4$ 
(dashed line), 8 (dot-dashed line), 16 (long-dashed line), and 32 (solid line).

\flushleft{Figure 6.}
The correlation time as a function of temperature
for the case $J=3$ and $\Delta =0$ and $N=4$
(dashed line), 8 (dot-dashed line), 16 (long-dashed line), and 32 (solid line).

\flushleft{Figure 7.}
The disclination density as a function of temperature
for the case $J=3$ and $\Delta =0$ and $N=4$
(dashed line), 8 (dot-dashed line), 16 (long-dashed line), and 32 (solid line).

\flushleft{Figure 8.}
The disclination asymmetry as a function of temperature
for the case $J=3$ and $\Delta =2$ and $N=4$
(dashed line), 8 (dot-dashed line), 16 (long-dashed line), and 32 (solid line).
The error bars are $\pm$ one standard deviation.

\flushleft{Figure 9.}
The disclination asymmetry as a function of temperature
for the case $J=3$ and $\Delta =3$ and $N=4$
(dashed line), 8 (dot-dashed line), 16 (long-dashed line), and 32 (solid line).
The error bars are $\pm$ one standard deviation.

\end{document}